\documentclass[11pt]{article}

\usepackage{enumitem}
\usepackage{epsfig}
\usepackage[mathscr]{euscript}
\usepackage{csquotes}

\usepackage[linesnumbered, ruled,noend,noline]{algorithm2e}

\usepackage{float}
\usepackage{graphicx}
\usepackage{ stmaryrd }
\usepackage{multirow}
\usepackage{gensymb}
\usepackage{array}
\usepackage{listings}
\usepackage{amsthm}

\renewcommand\thesection       {\arabic{section}}

\usepackage{amssymb}
\usepackage{graphicx,amsmath,calc}

\numberwithin{equation}{section}

\title{Information Redistribution Under Reductions in NP Search}
\author{Jing-Yuan Wei\thanks{Zhejiang Yi-Neng Grid-Storage Energy Co. Ltd. 
5277 Ouyang Rd, Haijing District, Wenzhou, Zhejiang, China. Email: weijingyuan@gmail.com}
}
\date{May 2026} 

\begin{document}
\maketitle

\begin{abstract}
Using reductions from structured P-matrix violation search to
classical NP-complete formulations such as 3-SAT and Subset Sum, we
examine the relationship between representational expansion,
auxiliary variables, local inferability, and information
accessibility. Rather than viewing reductions purely as computational
transformations, we interpret them as mechanisms that redistribute hidden witness
information across enlarged representations.

From this perspective, reductions, gadgets, and auxiliary structures
may expose globally encoded witness information to local propagation
and inference, while search algorithms act as decoding procedures
attempting to recover the original hidden witness.

The resulting observations suggest that representational expansion
may improve local inferability by introducing auxiliary variables and
consistency structures, while preserving the need to recover the
underlying witness information.

This work is exploratory in nature and proposes a conceptual framework
for understanding how reductions reshape information accessibility in
NP search.
\end{abstract}

\noindent\textbf{Keywords:}
NP search; P-matrix violation; information accessibility;
information theory; reductions; local inference; representational
expansion.

\section{Introduction}

The theory of NP-completeness traditionally studies reductions as
computational transformations between decision and search problems. A
reduction maps one representation into another while preserving
solvability, allowing apparently different problems to be studied
within a unified complexity-theoretic framework.

In this work we investigate a complementary perspective: reductions as
transformations of information accessibility.

Earlier work~\cite{wei2026-1} introduced an information-theoretic
viewpoint on structureless NP search under restricted interaction
models, while subsequent work~\cite{wei2026} extended this perspective
to structured search problems arising from sparse P-matrix violations.

Our starting point here is a structured search problem arising from
P-matrix violation detection. In the sparse-violation regime,
particularly when a matrix admits a unique violating principal minor,
the witness is globally encoded and difficult to localize through
local interactions. Prior work~\cite{wei2026} showed that under
restricted query models, each interaction reveals only negligible
information about the hidden violating subset.

This naturally raises the following question:

\begin{quote}
What happens to this information-accessibility barrier under reductions
to classical NP-complete problems?
\end{quote}

At first glance, reductions to problems such as 3-SAT appear to expose
much richer local structure. Clause propagation, conflict-driven
learning, and PPSZ-style inference procedures can aggregate
information through local interactions unavailable in the original
representation.

At the same time, reductions enlarge the representation through
auxiliary variables, slack variables, gadgets, and consistency
constraints. 
From an information-accessibility perspective, these enlarged
structures may be viewed as representational encodings that expose
globally encoded witness information to local propagation and
inference, while search algorithms operate as decoding procedures
attempting to recover the hidden witness through local interactions.

This viewpoint suggests a possible tradeoff. Reductions may improve
local accessibility by introducing auxiliary variables and consistency
structures that support propagation and inference, but such gains are
typically accompanied by representational expansion. The hidden
witness information is therefore not eliminated by the reduction, but
redistributed across a larger encoded structure through which it
becomes more locally accessible.

Symbolically, this may be viewed as
\[
\text{local accessibility}
\quad \leftrightarrow \quad
\text{representational expansion}.
\]

This work is exploratory in nature and presents illustrative
observations and conceptual interpretations suggesting that reductions
reshape information accessibility while preserving the need to recover
the original hidden witness.

Section~\ref{sc:pmatrix} reviews the information-accessibility barrier
in sparse P-matrix violation search.
Section~\ref{sc:reduction} interprets reductions as transformations of
information accessibility, while
Section~\ref{sc:observation} presents illustrative observations based
on reductions to 3-SAT and Subset Sum.
Section~\ref{sc:inference} discusses information accessibility and
SAT-style inference procedures such as PPSZ.
Finally, Section~\ref{sc:conclusion} concludes the paper.

\section{Sparse P-Matrix Violations and Accessibility Barriers}
\label{sc:pmatrix}

Consider a structured matrix family
\[
A(u,v)=B+uv^{\top},
\]
where \(B\) is an \(N\times N\) P-matrix and \(u,v\) are perturbation
vectors.

In sparse-violation regimes, suitable choices of perturbation vectors
\(u,v\) may yield a unique subset
\[
W\subseteq [N]
\]
such that
\[
\det(A(u,v)_W)\le 0,
\]
while the remaining principal minors remain positive.

The search problem is to identify the hidden violating subset \(W\).

In the oracle interaction model studied in~\cite{wei2026}, a query
tests whether a chosen subset equals the hidden witness. Under a
uniform prior over witnesses - motivated by the absence of strong
local inferability among principal minors in general matrix families
 - a fresh query succeeds with exponentially small probability:
\[
\Pr(Y=1)\approx 2^{-N}.
\]

Consequently, each interaction reveals only negligible information
about the hidden witness. From the information-theoretic viewpoint,
the witness carries
\[
H(W)\approx N
\]
bits of uncertainty, while the information gained per query vanishes
exponentially.

Earlier work~\cite{wei2026} showed that, despite strong underlying
algebraic structure, local observations may provide only weak
eliminative information about the hidden witness. Under restricted
interaction models, polynomially many interactions accumulate only
limited mutual information about \(W\).

This motivates the notion of an information-accessibility barrier in
structured search problems:

\begin{quote}
the relevant witness information exists globally, but remains only
weakly inferable through the available interactions.
\end{quote}

\section{Reductions as Information Redistribution}
\label{sc:reduction}

Traditional reductions preserve solvability. Here we view them through
a complementary lens: transformations of information accessibility and
local inferability.

We consider reductions from globally encoded search problems, such as
sparse P-matrix violation search, through standard NP-completeness
pipelines, for example
\[
\text{P-matrix violation}
\to
\text{3-SAT}
\to
\text{Subset Sum}.
\]

Two complementary effects arise under such reductions.

First, the reduced representations often expose substantially richer
local structure than the original formulation. For example,
reductions to 3-SAT expose clause interactions and propagation
procedures, while reductions to Subset Sum expose arithmetic
consistency relations and combinatorial merging structures.

Second, these transformations typically enlarge the representation
through auxiliary variables, slack variables, gadgets, and
consistency constraints. Suppose a reduction transforms an original
problem of dimension \(N\) into a representation of total dimension
\[
d=N+M,
\]
where \(M\) denotes the number of auxiliary variables or additional
representational coordinates introduced by the encoding. The quantity
\(M\) therefore measures the representational expansion introduced by
the reduction.

From the information-accessibility perspective, these auxiliary
structures may redistribute hidden witness information across enlarged
local structures that become accessible to propagation and inference
procedures. Clause propagation, PPSZ-style inference, and hierarchical
merging procedures can then aggregate information through repeated
local interactions unavailable in the original representation.

Nevertheless, the original hidden witness still carries
approximately
\[
H(W)\approx N
\]
bits of intrinsic information. After reduction, this information is
distributed across a larger representational space involving
auxiliary variables and consistency structures. This suggests the following interpretation:

\begin{quote}
reductions may redistribute information accessibility rather than
eliminate the original information requirement.
\end{quote}

Under this viewpoint,
\[
\text{reduction}
\approx
\text{encoding},
\]
while the associated search algorithm acts as a decoding procedure
attempting to recover the original hidden witness.

In this sense, the original representation may concentrate witness
information globally, while the reduced representation redistributes
that information across enlarged local structures and consistency
relations. The reduction therefore changes how the hidden information
becomes locally accessible without eliminating the need to recover the
original witness information.

\section{Illustrative Observations from 3-SAT and Subset Sum}
\label{sc:observation}

We illustrate the preceding viewpoint using a small \(N=6\) example
based on a unique-witness P-matrix violation instance. The reductions
considered here follow standard NP-completeness pipelines through
3-SAT.

\subsection{Dimensional Expansion}

\begin{table}[h!]
\centering
\footnotesize
\caption{\footnotesize{Illustrative dimensional expansion (\(N=6\)) under selected reduction encodings}}
\label{tab:expansion}
\vskip 5pt
\begin{tabular}{@{}lcccc@{}}
\hline
                 & Auxiliary    & Ratio & Total \\
Reduction Shape & Coordinates \((M)\) & \((M/N)\) & Dimension \((d)\) \\
\hline
Direct Matrix & 0 & 0.0 & 6 \\
3-SAT (Tseytin-style) & 12 & 2.0 & 18 \\
Subset Sum (via 3-SAT encoding) & 42 & 7.0 & 48 \\
\hline
\end{tabular}
\end{table}

The auxiliary-coordinate counts shown above depend on the chosen
reduction pipeline and encoding. The 3-SAT expansion is based on a
simplified Tseytin-style encoding~\cite{tseytin}, while the Subset
Sum dimensions are derived from standard NP-completeness reductions
through 3-SAT~\cite{cook1971,karp1972}. A detailed derivation of the
illustrative dimensional expansions is provided in
\ref{app:expansion}.

For simplicity, we use a solution-aware Tseytin-style reduction from
P-matrix violation to 3-SAT in which the unique violating subset is
already reflected in the construction. This yields a minimal
illustrative baseline with \(2N\) auxiliary variables. General
Tseytin-style circuit encodings, as well as full Cook-style
reductions, introduce substantially more auxiliary variables because
they must encode determinant computations and consistency relations
without prior knowledge of the violating subset.

\subsection{Representative Accessibility Comparison}

For the purpose of comparison, the determinant-computation stage may be
factored out, since any faithful reduction from P-matrix violation must
ultimately derive determinant-related conditions from the original
matrix instance. The comparison therefore focuses on the logical
accessibility layer induced by the reduction.

In the original psocid-style regime~\cite{wei2026-1,wei2026}, a single
query directly returns a binary yes/no answer associated with a chosen
principal minor. After reduction, however, algorithms such as PPSZ,
HGJ, and BBSS no longer access this information directly. Instead,
they attempt to recover it through global consistency search over a
large encoded Boolean system.

\begin{table}[h!]
\centering
\footnotesize
\caption{\footnotesize{Representative time-space accessibility comparison}}
\label{tab:time-space}
\vskip 5pt
\begin{tabular}{@{}lccc@{}}
\hline
                & Number of & Space & Accessibility \\
Reduction Shape & Trials & Complexity & Mechanism \\
\hline
Direct Matrix
& $O(2^N)$
& poly$(N)$
& direct witness search \\

3-SAT (PPSZ-style~\cite{ppsz1998,Hertli})
& $O(2^{0.386d})$
& poly$(d)$
& clause propagation \\

Subset Sum (HGJ-style~\cite{hgj})
& $O(2^{0.311d})$
& $O(2^{0.256d})$
& representation merging \\

Subset Sum (BBSS-style~\cite{BBSS20,li2025})
& $O(2^{0.24d})$
& $O(2^{0.222d})$
& hierarchical filtering \\
\hline
\end{tabular}
\end{table}

Under this viewpoint, one determinant query in the original regime may
be viewed as roughly analogous to one global search trial in the
reduced regime at the level of logical information access, despite the
substantial differences in representation and primitive operations.
After factoring out determinant computation, both the original witness
coordinates and the auxiliary variables introduced by the reduction are
treated uniformly as Boolean variables participating in a global
consistency structure.

This provides a coherent basis for comparing query complexity in the
original regime with trial complexity after reduction, while avoiding
ambiguities arising from representation-dependent arithmetic
operations. The comparison therefore concerns not primitive arithmetic
costs, but the accessibility of hidden witness information under
different representational encodings.

Table~\ref{tab:time-space} summarizes representative time-space
complexities for several algorithms for 3-SAT and Subset Sum.
Table~\ref{tab:complexity} combines these complexities with the
illustrative dimensional expansions in
Table~\ref{tab:expansion} to estimate the resulting search scales
after reduction.

\begin{table}[h!]
\centering
\footnotesize
\caption{\footnotesize{Illustrative complexity comparison (\(N=6\))}}
\label{tab:complexity}
\vskip 5pt
\begin{tabular}{@{}lccc@{}}
\hline
                & Dimension & Typical & Ratio to \(2^N\) \\
Reduction Shape & \((d)\) & Time Complexity & \((N=6)\) \\
\hline
Direct Matrix & 6
& $O(2^6)=O(64)$
& $\approx 1.0$ \\

3-SAT (PPSZ-style) & 18
& $O(2^{0.386d})\approx O(123)$
& $\approx 1.9$ \\

Subset Sum (HGJ-style) & 48
& $O(2^{0.311d})\approx O(3.1\times10^4)$
& $\approx 487$ \\

Subset Sum (BBSS-style) & 48
& $O(2^{0.24d})\approx O(2.9\times10^3)$
& $\approx 46$ \\
\hline
\end{tabular}
\end{table}

For the Subset Sum rows, we report time complexity only. The reduction
pipeline through 3-SAT further enlarges the dimensional expansion,
although direct reductions from P-matrix violation to Subset Sum may
introduce fewer auxiliary coordinates. Moreover, representation-based
Subset Sum algorithms typically use exponential space, so their
improved running times should be interpreted as part of a broader
time-space tradeoff.

Although the algorithms, reductions, and asymptotic constants differ
substantially across representations, a qualitative pattern appears
consistently across these examples:

\begin{itemize}
\item reductions enlarge the representational dimension;
\item auxiliary structures expose information to stronger local
propagation and inference procedures;
\item yet the original hidden witness information must still be
recovered through the enlarged representation.
\end{itemize}

\section{Information Accessibility and SAT-style Inference}
\label{sc:inference}

One possible interpretation of the preceding observations is that
reductions alter information accessibility by redistributing hidden
witness information across an enlarged representational space. From
the information-accessibility perspective, this suggests that
reductions may improve local inferability by redistributing hidden
witness information across larger encoded structures rather than
eliminating the original informational burden.

The original hidden witness still carries approximately
\[
H(W)\approx N
\]
bits of intrinsic information. After reduction, however, this
information becomes distributed across a larger collection of
variables, constraints, and combinatorial structures introduced by
the encoding.

This enlarged representation may expose information to local
propagation and inference procedures unavailable in the original
formulation. Clause propagation, critical-clause inference, and
hierarchical merging procedures can then aggregate information through
repeated local interactions.

In this sense:

\begin{itemize}
\item the original representation concentrates witness information in
a globally encoded structure;
\item the reduced representation redistributes this information across
local structures and consistency relations.
\end{itemize}

The reduction therefore changes how hidden witness information becomes
locally accessible.

Algorithms such as PPSZ illustrate this phenomenon particularly
clearly. Critical clauses expose locally accessible information that
can be aggregated through repeated inference steps. From the
information-accessibility perspective, this behavior is not
contradictory to the weak-inferability barrier observed in globally
encoded search problems. Rather, it suggests that reductions to
clause-based formulations may expose information through auxiliary
logical structure and local consistency relations absent in the
original representation.

At the same time, such accessibility gains may be accompanied by
representational expansion through auxiliary variables and encoding
structure. This suggests a broader tradeoff between local
inferability and representational expansion:
\[
\text{local inferability}
\quad \leftrightarrow \quad
\text{representational expansion}.
\]

From this viewpoint, reductions may preserve the need to recover the
original hidden witness while altering the mechanisms through which
that witness information becomes locally inferable.

This raises a broader question:

\begin{quote}
To what extent can reductions improve local inferability without
proportionally enlarging the representation?
\end{quote}

Understanding this tradeoff may provide a useful perspective on how
reductions reshape information accessibility across different
NP-complete representations.

\section{Concluding Remarks}
\label{sc:conclusion}

In this work we interpret reductions between NP search problems not
merely as computational transformations, but as mechanisms that
redistribute hidden witness information across enlarged
representations. From this viewpoint, auxiliary variables,
consistency gadgets, and enlarged combinatorial structures may improve
local inferability by exposing information to propagation and
inference procedures unavailable in the original formulation.

Under this interpretation, reductions act as representational
encodings, while algorithms such as PPSZ, HGJ, and BBSS operate as
inference or decoding procedures attempting to recover the original
hidden witness through repeated local interactions.

Interestingly, the reverse reduction direction also exhibits a related
accessibility transformation. In the reduction of 3-SAT to P-matrix
violation underlying the co-NP-completeness result of
Coxson~\cite{Coxson1994}, locally expressed Boolean constraints become
embedded into globally coupled determinant conditions. Thus,
reductions in both directions appear to redistribute witness
information across different accessibility structures rather than
eliminate the underlying informational burden.

The observations presented here suggest that improvements in local
inferability may be accompanied by representational expansion through
auxiliary variables and consistency structures. From this viewpoint,
reductions do not eliminate the need to recover the original hidden
witness, but reshape how that witness information becomes locally
accessible.

This viewpoint naturally connects the following components within a
unified information-accessibility perspective:

\begin{itemize}
\item globally encoded algebraic structure,
\item clause propagation,
\item local inference,
\item auxiliary variables,
\item representation-based algorithms,
\item and NP reductions.
\end{itemize}

This present work is exploratory in
nature and suggests that information accessibility may
provide a useful complementary viewpoint for understanding structured
search problems and NP reductions. 
The dimensional expansions and complexity comparisons discussed here
depend substantially on the chosen representations, encodings, and
algorithms, and should therefore be viewed as illustrative rather
than canonical properties of the underlying NP-complete problems.

Nevertheless, the observations suggest that reductions may alter not
only computational representation, but also the local accessibility
structure through which hidden witness information becomes locally
inferable. Understanding such transformations more formally may
provide an interesting direction for future research.

\appendix
\renewcommand{\thesection}{Appendix~\Alph{section}}

\section{Illustrative Dimensional Expansion}
\label{app:expansion}

We briefly explain the dimensional-expansion estimates used in the
illustrative observations in Table~\ref{tab:expansion}.

\paragraph{P-matrix violation to 3-SAT.}

Consider an original witness problem of dimension \(N\). In the
illustrative encoding considered here, the 3-SAT representation
introduces approximately \(2N\) auxiliary Boolean variables through a
simplified Tseytin-style consistency construction. The resulting 3-SAT
instance therefore contains approximately
\[
d \approx 3N
\]
Boolean variables in total.

This \(3N\) estimate is intended only as an illustrative baseline. It
corresponds to a simplified witness-aware encoding used to estimate
representational expansion. More general Cook/Tseytin-style reductions,
which must encode determinant computations and consistency relations
without prior knowledge of the violating subset, may introduce
substantially more auxiliary variables.

\paragraph{3-SAT to Subset Sum.}

In the standard reduction from 3-SAT to Subset Sum, one introduces
\[
2n+2m
\]
Subset Sum items, where \(n\) denotes the number of Boolean variables
and \(m\) the number of clauses in the 3-SAT instance.

Assuming a representative sparse regime with
\[
m \approx N,
\]
and substituting \(n \approx 3N\), we obtain
\[
2n+2m
=
2(3N)+2N
=
8N.
\]

Thus the resulting Subset Sum representation contains approximately
\(8N\) coordinates/items in total. Relative to the original witness
dimension \(N\), the auxiliary expansion is therefore
\[
M = 8N - N = 7N.
\]

For the illustrative \(N=6\) example, this yields
\[
M=42,
\qquad
d=48,
\]
as shown in Table~\ref{tab:expansion}.

The \(3N\) and \(8N\) estimates discussed above serve different
illustrative purposes and should not be interpreted as canonical or
directly comparable reduction bounds. The \(3N\) estimate corresponds
to a simplified witness-aware 3-SAT encoding, whereas the \(8N\)
estimate follows a standard NP-completeness reduction pipeline through
3-SAT intended to preserve the full feasible solution
structure~\cite{cook1971,karp1972}.

More generally, the dimensional expansions discussed in this work
depend substantially on the chosen reduction pipeline and encoding, and
are intended only as representative illustrations of the
representational expansion introduced by standard reductions. In
particular, standard 3-SAT-to-Subset-Sum reductions may introduce
additional filler coordinates to manage carries in positional
encodings. However, for coarse asymptotic dimensional estimates, the
dominant contribution typically remains of the form \(2n+2m\).


\end{document}